\documentclass[prl,superscriptaddress,twocolumn,showpacs,preprintnumbers,amsmath,floatfix]{revtex4}
\usepackage{amssymb}
\usepackage{graphicx}
\usepackage{color}

\begin{document}

\title{Fully Spin-Polarized Current in Gated Bilayer Silicene}
\author{Xiao-Fang Ouyang}
\affiliation{Shanghai Key Laboratory of Special Artificial Microstructure Materials and Technology, School of Physics Science and engineering, Tongji University, Shanghai 200092, P.R. China}
\affiliation{School of new energy science and engineering, Xinyu University, Jiangxi 338000, P.R. China}
\author{Ze-Yi Song}
\affiliation{Shanghai Key Laboratory of Special Artificial Microstructure Materials and Technology, School of Physics Science and engineering, Tongji University, Shanghai 200092, P.R. China}
\author{Yu-Zhong Zhang}
\email[Corresponding author.]{Email: yzzhang@tongji.edu.cn}
\affiliation{Shanghai Key Laboratory of Special Artificial Microstructure Materials and Technology, School of Physics Science and engineering, Tongji University, Shanghai 200092, P.R. China}
\date{\today}

\begin{abstract}
\centerline{Abstract}
By applying density functional theory calculations, we predict that the groundstate of bilayer silicene at certain interlayer distances can be antiferromagnetic. At small electron or hole doping, it becomes half metallic under applied out-of-plane electric field, which can be used to produce fully spin-polarized field-effect-driven current even in the absence of external magnetic field, ferromagnetic substrates, doped magnetic ions, or spin-orbital coupling. Our finding points out a new route to overcome the major challenge of spintronics.
\end{abstract}

\maketitle

Spintronics, where intrinsic spin degrees of freedom of electrons are controlled and manipulated for building logic and memory devices, has been an active research field for decades~\cite{WolfScience2001}. A central challenge in this field is the difficulty to generate fully spin-polarized current. Half metals, where only one spin channel is conductive while the other is insulating~\cite{GrootPRL1983}, can resolve this problem, which leads to numerous theoretical predictions and experimental investigations on half metallic ferromagnets or antiferromagnets with transition-metal ions~\cite{HuangPRB2003,YuScience2001,HiroeJSAP2000,ChioncelPRB2003}. However, these systems with metallic ions may not be easily integrated with silicon-based semiconductors due to the interface, low solubility of magnetic ions in semiconductors~\cite{KazakovaPRB2008}, or short spin relaxation time~\cite{KrainovPRB2017}, which calls for searching metal-free materials with half-metallicity.

Since the discovery of graphene which exhibit extraordinary thermal, mechanical, and electrical properties~\cite{NovoselovScience2004,MorozovPRL2008}, many two-dimensional carbon-based materials with half metallicity have been theoretically proposed, including graphene nanoribbons with applied in-plane electric field~\cite{ShengPRB2014,SonNature2006} or different chemical modifications of the edge~\cite{ZhengPRB2008}, semihydrogenations of graphene~\cite{ZhouNanoletter2009,YangSciencereport2014,LiAPL2011}, and graphitic carbon nitride~\cite{DuPRL2012}, etc. In spite of extensive design based on first principles calculations, half metallicity has not yet been experimentally confirmed in all the predicted two-dimensional carbon-based materials.

Silicene, a silicon counterpart of graphene, is a particularly promising material for silicon-based technology~\cite{VerriPRB2007,CiraciPRL2009,WuNanolett2012,VogtPRL2012}. Due to the resemblance between silicene and graphene, similar ideas of realizing half metallicity as was done in graphene have also been theoretically carried out for silicene. However, the predicted half metallicity requires either a strong inplane electric field~\cite{WANGNano2012} or an edge modification with atomic precision~\cite{PanPhysicE2014,YangJAP2014}, which makes experimental synthesis quite difficult. Recently, a new design based on gated silicene was proposed, where two nearly 100\% spin-polarized Dirac cone with opposite polarization appears due to the presence of both spin-orbital coupling and spatial inversion symmetry breaking induced by out-of-plane electric field on the buckled rather than flat structure~\cite{Tsainaturecommunication2013}. However, a sizeable exchange field induced by either ferromagnetic substrate/adatoms or a magnetic field is needed to filter spins with specified spin direction. Therefore, a natural question arises; is there other feasible way to generate fully spin-polarized current based on silicene?

In this letter, by taking the advantage of buckled structure of silicene, we would like to propose that, even without ferromagnetic substrate/magnetic field/doping magnetic ions, cutting silicene into nanoribbon, or selective modification like semihydrogenations, and at the same time, in the absence of spin-orbital coupling, silicene under an out-of-plane electric field could be a candidate for a spintronics device to generate fully spin-polarized current, as long as silicene were in an antiferromagnetic state. The underlying physics can be understood by following tight-binding model.

\begin{eqnarray}
&H&=\sum_{k \sigma} (\varepsilon(k) c^{\dagger}_{k A \sigma} c_{k B \sigma}+h.c.)\label{eq:hamiltonian} \\
&+& V\sum_{k \sigma} (n_{k A \sigma}-n_{k B \sigma})+\sum_{k \sigma} \sigma M (n_{k A \sigma}-n_{k B \sigma}), \nonumber
\end{eqnarray}
where $c^{\dagger}_{k\gamma\sigma}$ ($c_{i\gamma\sigma}$) creates (annihilates) an electron in subblatice $\gamma=A$ and $B$ of momentum $k$ with spin $\sigma$ and $n_{k\gamma\sigma}=c^{\dagger}_{k\gamma\sigma}c_{k\gamma\sigma}$ is the occupation operator. The first term describes the band structure of silicene while the second and the third terms represent for a staggered potential induced by applied out-of-plane electric field on silicene with buckled structure and a staggered magnetic field induced by either strong onsite Coulomb interaction or the Fermi surface nesting, respectively.

The Hamiltonian (\ref{eq:hamiltonian}) can be easily diagonalized and the eigenvalues are given by
 \begin{equation}\label{bands}
 \epsilon^{\sigma}_{1,2}(k)=\pm\sqrt{(V+\sigma M)^2+|\varepsilon(k)|^2}.
 \end{equation}

\begin{figure}[htbp]
\includegraphics[width=0.48\textwidth]{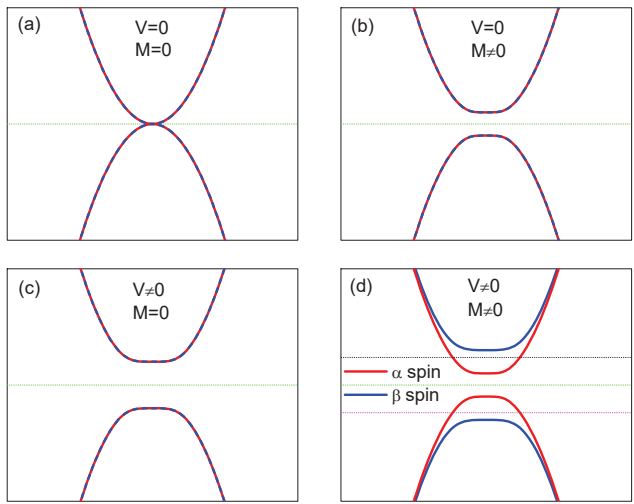}
\caption{(Color online) Effects of staggered potential induced by out-of-plane electric field on silicene with buckled structure and staggered magnetic field caused by strong electronic interaction on the bands close to the Fermi level. Two different spin species $\alpha$ and $\beta$ are shown in red and blue, respectively. Green horizontal dotted line denotes the Fermi level in half-filled case, while the black and magenta ones in (d) the Fermi levels after small electron and hole doping,respectively.}
\label{Fig:one}
\end{figure}

Fig.~\ref{Fig:one} illustrates change of bands close to the Fermi level in the presence or the absence of staggered potential $V$ and staggered magnetic field $M$. Without $V$ and $M$, different spin states are degenerate and the system is metallic as shown in Fig.~\ref{Fig:one} (a). When $V$ or $M$ is separately turned on, a gap appears while degeneracy remains in spin states in each case as seen in Fig.~\ref{Fig:one} (b) and (c). When both $V$ and $M$ are taken into account, the spin degeneracy is lifted near the Fermi level. Then, both small electron and hole doping will shift the Fermi level into the bands with fully polarized spin in one direction as displayed in Fig.~\ref{Fig:one} (d), resulting in a half metal realized on gated silicene. This half-metallicity can be simply controlled by external out-of-plane electric field. Now the key question is whether there exists antiferromagnetic silicene under gate field which may satisfy above conditions of a staggered magnetic field with a staggered electric potential.

In the following, we will show based on density functional theory calculations that ground state of bilayer silicene can be antiferromagnetic when the interlayer distance exceeds a critical value of $2.6$\AA. By applied out-of-plane electric field, half metallicity occurs at small electron or hole doping. Our finding opens a new avenue towards searching new half metals compatible with present semiconductor technology.

Our investigations of gated bilayer silicene are based on projector augmented wave method\cite{BlochlPRB1994} as implemented in the Vienna Ab-initio Simulation Package(VASP)\cite{KressePRB1996,KresseCMS1996}. The generalized gradient approximation (GGA)\cite{PerdewPRL1996} for the exchange-correlation functional are employed with plane wave energy cutoff of $500$~eV. Brillouin-zone sampling is performed on $50\times50\times1$ k mesh determined by Monkhorst-Pack scheme. When the density of states is calculated, a $60\times60\times1$ Monkhorst-Pack k-points grid is used. To eliminate the interaction between two bilayer silicene, a sufficiently large vacuum of more than $15$~\AA~was selected. In order to verify that the designed structure is dynamically stable, we employ VASP in combination with phonopy\cite{TogoScripta2015,Phonopy2015} to calculate the phonon dispersion. Finally, we do molecular dynamics calculations to figure out the stability of structure and magnetism at finite temperature of $100$~K, which is well above the temperature of liquid nitrogen.

\begin{figure}[htbp]
\includegraphics[width=0.48\textwidth]{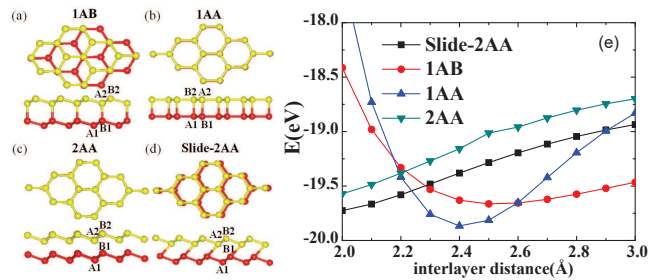}
\caption{(Color online) (a)-(d) are the cartoons for four stable structures of bilayer silicene labelled by 1AB, 1AA, 2AA, and Slide-2AA. For each structure, top and side views are shown. Four inequivalent carbon atoms in one unit cell are tagged by A1, B1 and A2, B2, respectively. (e) The total energies of four structures as a function of interlayer distance which is defined as z-component of the length between nearest neighbor atoms belonging to different layers.}
\label{Fig:two}
\end{figure}

In Fig.~\ref{Fig:two}, we present calculated total energies without considering magnetism as a function of interlayer distance which is defined as z-component of the length between nearest neighbor silicon atoms belonging to different layers based on four stable structures as shown in Fig.~\ref{Fig:two} (a)-(d) which is reported previously\cite{MengshengAPL2014}. Here, z-component denotes the direction perpendicular to the silicene plane. It is found that the ground state is dependent on the interlayer distance as can be seen in Fig.~\ref{Fig:two} (e). When the distance is larger than 2.2~\AA~but smaller than 2.6~\AA, 1AA configuration is the ground state where silicon atoms of one layer are just on top of those of the other and both layers are flat. As the distance is not larger than 2.2~\AA, slide-2AA configuration becomes the ground state where both layers are buckled and shifted slightly against each other. Once the distance is larger than 2.6~\AA, the 1AB configuration arises as the most stable solution where both layers remain buckled with one set of silicon atoms of one layer sitting above those of the other layer and the other set of silicon atoms located in the center of the hexagon of the other layer. The 2AA configuration will never be the ground state for any interlayer distance where both layers are buckled without shift.

\begin{figure}[htbp]
\includegraphics[width=0.48\textwidth]{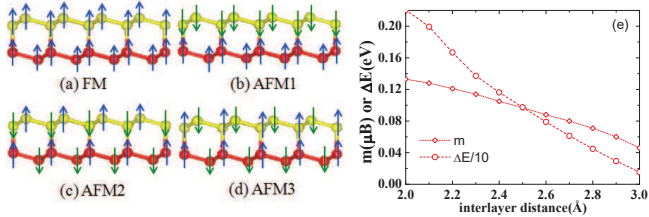}
\caption{(Color online) (a)-(d) are cartoons for four spin states on 1AB structure labelled by FM, AFM1, AFM2, and AFM3. Only AFM2 with both intralayer and interlayer antiferromagnetism is a stable solution based on spin-polarized GGA calculations. (e) The total energy difference between nonmagnetic (NM) and AFM2 states given by $\Delta E= E_{NM}-E_{AFM2}$ and the magnetization of AFM2 state on A1 or B2 site (see the definition in Fig.~\ref{Fig:two} (a)) as a function of interlayer distance.}
\label{Fig:three}
\end{figure}

After figuring out the structural configurations as a function of interlayer distance, we perform spin polarized GGA calculations to search possible existence of magnetic states. Four different spin states are taken into account, including one ferromagnetic (FM) state, and three antiferromagnetic (AFM) states as illustrated in Fig.~\ref{Fig:three} (a)-(d). While AFM1 is ferromagnetic within each layer but antiferromagnetic between layers, AFM3 is intralayer antiferromagnetic with interlayer ferromagnetic. AFM2 is the state with both intralayer and interlayer antiferromagnetic arrangements.

We find that all the above mentioned magnetic states are unstable on 1AA, 2AA, and slide-2AA configurations where all the spin polarized GGA calculations are converged to nonmagnetic solutions. Only 1AB structure has a stable solution of AFM2 state. The energy difference between nonmagnetic and AFM2 state is shown in Fig.~\ref{Fig:three} (e). It is found that the AFM2 state always has lower total energy than nonmagnetic state at all the interlayer distances we studied. However, the difference becomes smaller as interlayer distance becomes larger. This is due to the fact that the magnetic moment on silicon atom is monotonously reduced as a function of interlayer distance as shown in Fig.~\ref{Fig:three} (e). Since magnetic moment vanishes on monolayer silicene and magnetic solution does not exist on other structures, we can conclude that the interlayer couplings between silicon atoms centered in the hexagon play important roles in the formation of magnetic state. In the following, we will focus on the bilayer silicene with interlayer distance of $2.7$~\AA~where ground state is AFM2.

\begin{figure}[htbp]
\includegraphics[width=0.48\textwidth]{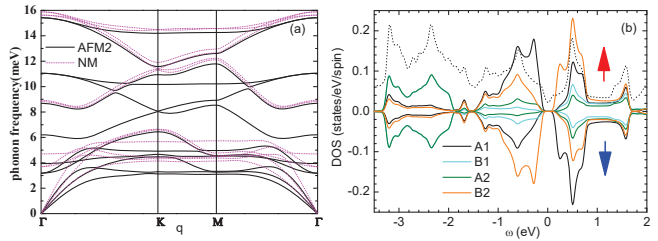}
\caption{(Color online) (a) Phonon dispersion of bilayer silicene with 1AB structure in nonmagnetic (NM) and AFM2 states. (b) atomic resolved partial densities of states of p$_z$ orbitals of bilayer silicene with 1AB structure in AFM2 state. Total density of states of NM states divided by a factor of $20$ is shown by dotted line.}
\label{Fig:four}
\end{figure}

In Fig.~\ref{Fig:four} (a), phonon spectrum are shown for bilayer silicene with 1AB structure. It is found that all the phonon frequencies are positive for both nonmagnetic and AFM2 states, indicating that bilayer silicene with 1AB structure are dynamically stable. There are 3~acoustic  and 9~optical branches in total due to 4~silicon atoms per unit cell. Around 4~meV, the 3~optical branches are mainly contributed from out-of-plane vibrations, while above 6~meV, the in-plane vibrations become dominant. We find strong softening and splitting of phonons due to the appearance of magnetism by comparing the phonon spectrum of AFM2 state with that of nonmagnetic state.

Fig.~\ref{Fig:four} (b) only shows the atomic resolved partial density of state (DOS) of $p_z$ orbitals for bilayer silicene with 1AB structure, since the effects of other orbitals on the magnetic moment and low-energy electronic states are negligibly small. Compared to nonmagnetic state where finite DOS is present at the Fermi level, a gap opens due to the occurrence of magnetic order in AFM2 state. In the vicinity of the band gap, the DOS are largely contributed from $p_z$ orbital of A1 and B2 silicon atoms, which is consistent with the fact that the calculated magnetic moment mainly exists on A1 and B2 sites. The small magnitude of magnetization on A2 and B1 site (see Fig.~\ref{Fig:two} (a) for the definition of different sites) may be ascribed to strong hybridization between $p_z$ orbitals on these two sites which are just on top of each other, since strong hybridization generally favors delocalization of electrons and suppresses formation of local spin. Moreover, we find that majority and minority spins on A1 and B2 sites are of opposite directions and degenerate, respectively. Therefore, it is expectable that with further applied electric field perpendicular to the silicon plane, A1 and B2 sites will have different electric potential energies due to their different positions along the direction of electric field. Consequently, above mentioned conditions of a staggered magnetic field with a staggered electric potential are fulfilled, and spin degeneracy between A1 and B2 sites should be lifted.

\begin{figure}[htbp]
\includegraphics[width=0.48\textwidth]{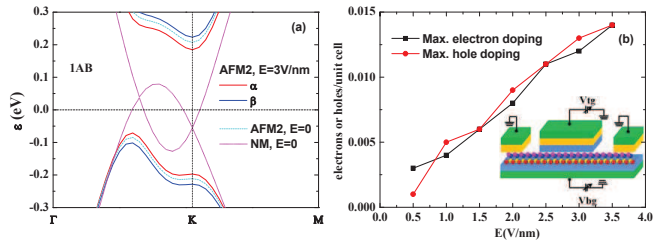}
\caption{(Color online) (a) Band structure close to the Fermi level for bilayer silicene with 1AB structure along the path of $\Gamma-K-M$ for three cases; nonmagnetic and AFM2 states without gate field, and AFM2 state with gate field of $E=3$~V/nm where spin degeneracy is lifted into two species of $\alpha$ and $\beta$. (b) Upper limit of electron or hole doping, under which doped bilayer silicene with gate field can be a half metal. The inset exhibits a cartoon for a dual-gate field effect transistor device which can realize the function of 100\% spin filter or spin generator.}
\label{Fig:five}
\end{figure}

The expected phenomenon indeed takes place as we apply out-of-plane electric field to the bilayer silicene in our first principles calculations. In Fig.~\ref{Fig:five} (a), we present band structure along the path of $\Gamma-K-M$ for nonmagnetic case as well as AFM2 cases with and without electric field. In the absence of magnetic order and electric field, the bilayer silicene is metallic with two bands crossing the Fermi level. As AFM2 order is taken into account, an indirect gap appears while spins remain degenerate. After external electric field of $3$~V/nm is turned on, the bands of opposite spin states split into two nondegenerate $\alpha$ and $\beta$ bands as can be seen in Fig.~\ref{Fig:five} (a). If slightly doping electrons or holes into such a bilayer silicene system with the Fermi level only crossing $\alpha$ band, one can obtain an antiferromagnetic half metal where only electrons with $\alpha$ spin species is conductive while the electrons with opposite spin are insulating. The underlying mechanism for this half metallicity is exactly the same as described in Fig.~\ref{Fig:one}.

Such a 100\% spin filter or spin generator device can be possibly realized by a dual-gate bilayer silicene field effect transistor device as illustrated in the inset of Fig.~\ref{Fig:five} (b). By using both top and bottom gates in the bilayer silicene field effect transistor device, one is able to control independently the splitting of electronic bands with opposite spins and the carrier doping concentration. However, in order to obtain fully spin-polarized current, the electron or hole doping has an upper limit, above which the Fermi level will enter the $\beta$ band and as a consequence electrons with opposite spin also become conductive. Therefore, we calculate maximum electron or hole doping as a function of gate field within rigid band approximation as displayed in Fig.~\ref{Fig:five} (b). The maximum number of electrons or holes which can be doped into gated antiferromagnetic bilayer silicene without destroying half metallicity increases as the gate field increases. This is due to the fact that the splitting of bands with opposite spins becomes larger at larger  gate field.

Finally, we investigate the stabilities of bilayer silicene with 1AB structure and AFM2 order at finite temperature of $100$~K which is well above the temperature of liquid nitrogen by {\it ab} initio molecular dynamics. The spin polarized GGA calculations are based on a $6\times6$ supercell with a Nos\'{e}-Hoover thermostat\cite{EvansJCP1985} of $100$~K. After $10$~ps, we find no structural and magnetic destruction of bilayer silicene. The robustness of lattice structure and the stability of AFM2 state indicate that there may be a well defined energy barrier between AFM2 state on 1AB structure and all the other configurations. On the other hand, we find negligible effect of spin-orbital coupling on the electronic properties of antiferromagnetic bilayer silicene with 1AB configuration.

In conclusion, we predict from density functional theory calculations that bilayer silicene can be antiferromagnetic and with out-of-plane gate field it can be a half metal under small electron or hole doping. This half metallicity does not require strong inplane electric field, strong spin-orbital coupling, high magnetic field or ferromagnetic substrate, doped magnetic ions, and modification of the edges or site-selective chemical reaction. In contrast, the half metallicity can be simply controlled by gated electrical field. Our findings indicate that there may exist a large amount of two-dimensional materials with half metallicity controlled by gated electric field, as long as the candidates can be antiferromagnetic and the antiferromagnetically arranged atoms are not located in the same plane.

\section{Acknowledgement}\label{Acknowledgement}

This work is supported by National Natural Science Foundation of China (Nos. 11774258, 11174219) and Science and technology project in jiangxi province department of education (Grant No. GJJ161196).

\end{document}